\magnification=\magstep1
\overfullrule=0pt
\hsize=6true in
\hoffset=0.45true in
\baselineskip=16pt plus 2pt
\parskip=7pt
\raggedbottom

\font\third=cmr10 scaled\magstep0
\font\first=cmr7 at 12truept
\centerline{\bf FLUID MODELS FOR KINETIC EFFECTS ON COHERENT}
\centerline{\bf NONLINEAR ALFV\'EN WAVES  I: FUNDAMENTAL 
THEORY}
\vskip 0.3 cm
\centerline{(published: {\sl Phys. Plasmas}, {\bf 3}, 863 (1996) )} 
%
\sl
\vskip .7 cm
\centerline{{\rm M.V.~Medvedev}\footnote\dag{{\rm Also:} Russian 
Research
Centre ``Kurchatov Institute", Institute for Nuclear Fusion, Moscow 123182,
RUSSIA.}{\rm \  and P.H.~Diamond}\footnote\ddag{{\rm Also:} 
General Atomics, San Diego, California 92122.} }

\centerline{Physics Department, University of California at San Diego}
\centerline{La Jolla, California 92093-0319}   
\vskip .3 cm
\centerline{\bf Abstract}
\vskip 0.1 cm
\third
Collisionless regime kinetic models for coherent nonlinear Alfv\'en wave 
dynamics
are studied using fluid moment equations with an approximate closure 
anzatz. Resonant particle effects are modelled by incorporating an 
additional term representing dissipation akin to parallel heat conduction. 
Unlike collisional
dissipation, parallel heat conduction is presented by an integral operator. 
The modified derivative nonlinear Schr\"odinger equation thus has a spatially
nonlocal nonlinear term describing the long-time evolution of the envelope of
parallel-propagating Alfv\'en waves, as well. Coefficients  in the 
nonlinear terms are free of the  $(1-\beta)^{-1}$ singularity
usually encountered in previous analyses, and have very a
simple form which  clarifies the physical processes governing the large
amplitude Alfv\'enic nonlinear dynamics.
The nonlinearity appears via coupling of an Alfv\'enic mode to a 
kinetic ion-acoustic mode. Damping
of the nonlinear  Alfv\'en wave appears via strong Landau damping
of the ion-acoustic wave when the electron-to-ion temperature ratio is close
to unity. For a (slightly) obliquely propagating wave, there are finite Larmor 
radius corrections in the dynamical equation. 
This effect depends on the angle of wave 
propagation relative to $B_0$ and vanishes for the limit of strictly parallel 
propagation. Explicit magnetic perturbation envelope equations amenable to
further analysis and numerical solution are obtained.
Implications of these models for collisionless shock dynamics
are discussed.
\vskip 0.2 cm
\line{PACS numbers: 52.35.Mw, 52.30.-q, 96.50.Ek \hfil\break}
\vfill
\eject
\first
\vskip 1.0cm
{\bf I. Introduction }
\vskip 0.5cm
It is widely believed  that Alfv\'en waves play an important role in 
interplanetary  plasmas. 
High level of magnetohydrodynamic (MHD) wave activity and shock
waves are observed in the planetary and solar wind plasmas [1-3].
The plasma environment is characterized by a quite  weak mean 
interplanetary magnetic field and large amplitude magnetic field fluctuations
caused by Alfv\'en and magnetosonic waves. Thus, nonlinear effects play an
important role in the evolution of such waves. Interest in large-amplitude
Alfv\'en wave dynamics arose from the attempts to understand
the steepening of wave trains and shock formation in space as well as from 
general interest in nonlinear waves as a whole.

Previous studies have shown [4-6] that nonlinear Alfv\'en and 
fast magnetosonic waves
are described by the derivative nonlinear Schr\"odinger equation (DNLS).
In this model, a parallel ponderomotive force of a high-amplitude wave
perturbs the plasma density. The force
squeezes plasma out the  
regions of larger magnetic field, locally decreasing the plasma density, and
increasing the Alfv\'en wave velocity.  
As the amplitude of  a circularly polarized Alfv\'en wave varies on 
time scales much slower than the wave frequency,  the section of a wave train
with higher amplitude propagates faster than the part with lower one. This
gives rise to nonlinear steepening of the large-amplitude Alfv\'en wave.
Dispersion, as usual, ultimately controls steepening. The DNLS thus
describes the long time-scale dynamics of the envelope of such Alfv\'en
waves.
The DNLS is an integrable equation, which describes soliton and
multi-soliton solutions, shock waves, modulation instability of solitons
[7-9] etc. As was shown by Longtin and Sonnerup [10] and Wong and
Goldstein [11] (from analyses of the full set of MHD equations),
modulation instability occurvs for left-hand polarized  waves if $\beta <1$
and for right-hand polarized waves if $\beta >1$. This DNLS
modulation instability is strongly sensitive to the sign of the coefficient of
the cubic nonlinearity, which is propotional to $(1-\beta)^{-1}$. Indeed, at
$\beta\cong 1$ (typical of solar wind plasma), the sound speed approaches 
the Alfv\'en
velocity, and resonant energy exchange between sound and Alfv\'en waves 
is strong, in the absence of parallel dissipation. Then, the Alfv\'en
wave is always in phase with the acoustic wave, giving rise to fast steepening
of the front of the nonlinear Alfv\'en wave train. The rate of the steepening is
roughly proportional to $(1-\beta)^{-1}$. Significant conversion of Alfv\'en
wave energy into ion thermal energy (i.e. ion heating) occurs as well.
It is interesting to note that the overwhelming preponderance of theoretical
work in this field of quasi-parallel Alfv\'enic shocks and solitons is based
upon a simple MHD plasma model. Thus, with the notable exceptions of
Ref.~[6,12-14], the theoretical `lore' of nonlinear Alfv\'en waves in a
 collisionless,
$\beta\sim1$ plasma is built upon a conceptual paradigm constructed for a
collisional, $\beta<1$ system. This state of affairs is due, in part, to the
intractability and unwieldiness of straightforward kinetic analysis of
nonlinear Alfv\'en waves. In this paper, we offer a simplier approach which
exploits recent developments in the theory of fluid modelling of kinetic effects.

We can expect that the influence of kinetic effects, such as  Landau
damping must become important when $\beta$ approaches unity. A fully 
kinetic calculation was undertaken by Rogister [5] 
for the case of a high-beta plasma. For
parallel propagation, his results counside with equations
obtained later by Mj\o lhus and Wyller [12, 6] and Spangler [13, 14]. The
effect of Landau damping appears in the DNLS via an additional cubic
term which is an integral operator over space. There is no 
$(1-\beta)^{-1}$ singularity in the nonlinear derivative term.
 The nonlocality is a consequence of the finite time history of an ion transit
through an envelope modulation (i.e. the coherence time of an ion with the
modulation is not infinitesimal in comparison to the modulation growth time).
Thus nonlinear Alfv\'en wave dynamics are nonlocal, and not governed by
{\sl local} spatial derivatives of the perturbed field, alone.
The other attempts to include Landau damping in the nonlinear Alfv\'en
wave evolution used kinetics to calculate second-order density (or pressure) 
perturbation, but the DNLS-like equation (for $\widetilde{\bf b}$) was
obtained from MHD equations. Mj\o lhus and Wyller [12, 6] used a
guiding center formalism and Spangler [13, 14] used the full Vlasov
equation. Both methods predict a nonlocal integral term, and the coefficients 
of the nonlinear terms  coinside with those of Rogister [5]. 
In these cases, the coefficient of the derivative cubic
nonlinear term does not change sign at $\beta =1$, except for large
values of $T_e/T_i$  (the electron-to-ion temperature ratio). 
The coefficient of the nonlocal term is always negative. Both
coefficients of the nonlinear terms (cubic and nonlocal) depend strongly on
$\beta$ and on $T_e/T_i$.

Obviously, a simple fluid model of collisionless shocks is desirable for
reasons of tractibility. However, kinetic effects are essential for describing
interplanetary plasma dynamics.
It is known that fluid models poorly represent most kinetic effects. Several
authors have suggested that kinetic effects, such as Landau damping may be
modelled in fluid equations by adding parallel dissipation terms [15,16]. 
Recently,
a closure method for modelling of kinetic effects was developed [17-21] and 
applied to ion-temperature gradient instabilities using gyrofluid models.   
This closure anzatz for fluid moment equations i) ensures particle,
momentum and energy conservation, ii) takes a simple form in
wave-number space, and iii) has a linear response function very close to that 
of a collisionless, Maxwellian plasma. In this paper we use the simplest
version of the method, namely that proposed by Hammet and Perkins [17].

In this paper we investigate effects of dissipation on the nonlinear, 
parallel-propagating Alfv\'en wave. We first use one-fluid MHD equations 
with parallel dissipation coefficients
 $\mu_\|$ and $\chi_\|$  which are constant and  independent
of mode frequency and wave-number. Later we replace
constant $\chi_\|$  with the integral operator representation of Hammet and 
Perkins [17], and obtain a modifield DNLS similar to that obtained by
Mj\o lhus and Wyller [12, 6] and Spangler [13, 14]. However, our approach
yields expressions for  coefficients of the nonlinear terms which are much 
simpler than theirs and allows clear, unambiguous physical interpretation of
the results.  
We also derive the  $T_e/T_i$ dependence using two-fluid MHD
equations. In an ion-acoustic wave, ion density perturbations of the electron 
background are mediated by electric field effects. 
In collisionless plasmas the ion-acoustic branch, thus, 
replaces  the acoustic branch as caused by gas-kinetic pressure perturbations.
The results indicate that the nonlinearity of Alfv\'en waves in collisionless 
plasma is controlled by the coupling of the Alfv\'en mode to 
an ion-acoustic mode, i.e. an ion-density perturbation mediated by an 
electron responce electric field. This is
in contrast to the conventional view [4,7,8] of this nonlinearity
as due to a ponderomotive (i.e. a gas-kinetic/collisional) 
plasma density perturbation. The kinetic
damping on resonant particles is nothing more than the
usual strong Landau damping of an ion-acoustic wave for
$T_e\simeq T_i$. This damping leads to enhanced ion heating which further
raises $T_i/T_e$.

We also  consider the case of a slightly obliquely propagating wave. There, 
one can expect that other kinetic effects may be relevant to the dynamics.
It is shown that the  ${\bf E}\times {\bf B}$ 
drift in the electric field produced by charge separation in an ion-acoustic 
wave with an ambient magnetic field is not significant. Another effect 
which is important, however, 
is caused by gyro-averaged electric and magnetic fields acting on a
particle over the scale of a Larmor orbit. (There are no $(k_\bot\rho_i)^2$ 
effects on dispersion in the DNLS approximation. Such effects may appear 
in higher-order in $\widetilde b/B_0$ calculations.)
This effect enters the modified DNLS 
in a way similar to collisional dissipation. The finite Larmor radius
correction depends on the angle of propagation of a wave and disappears for
a strictly parallel-propagating waves.
We explore  modulational  stability for the general case of a dissipative
nonlinear Alfv\'en wave. This instability is important when one considers the
origin and evolution of solitons, wave packets, shock waves, etc. 

The rest of this paper is organized as follows. In Section II we derive the
evolution equation for dissipative nonlinear parallel-propagating Alfv\'en
waves. In Section III the modified DNLS with the resonant particle effect is
obtained. In Section IV we consider the influence of finite Larmor radius
corrections to the dynamics of a slightly oblique, nonlinear Alfv\'en wave.
In Section V we investigate the modulation instability of dissipative nonlinear
Alfv\'en waves. 
Section VI is a discussion of the results obtained.
\vskip 1.0 cm
{\bf II. Dissipative Nonlinear Alfv\'en Waves}
\vskip 0.5 cm
We obtain the equation governing nonlinear wave dynamics from a multiple 
time scale
expansion of the dissipative MHD equations. Dissipation is included via 
the parallel viscosity $\mu_\|$ or the parallel heat conductivity 
$\chi_\|$. We considered  both these cases for the following reason.
The first case, $\mu\not=0$, corresponds to the three-moment fluid model 
of Landau damping. The four-moment fluid model should 
contain both $\chi_\|$ and $\mu_\|$ to close the equations. As shown in 
Ref.~[17], the best fit of the linear-response function of of this model to the  
linear-response function of a Maxwellian plasma is 
achieved when $\mu_\|=0$ and $\chi_\|\not=0$. This is the second 
case we consider here. Note that any closure of this sort which tacitly assumes
a Maxwellian plasma intrinsically fails to capture the strong local modification
of the distribution function associated with large-amplitude turbulence,
trapping, etc.

In the derivation, we follow Ref.~[4]. We assume a plane wave propagating
in the $z$ direction. Thus, all quantities are functions of $z$ and $t$. The MHD
equations are written as:
$$
\eqalignno{
&{\partial\rho\over\partial t}+\rho{\partial u\over\partial z}+u{\partial\rho
\over\partial z}=0 , &(1)\cr
&\rho{\partial u\over\partial t}+\rho u{\partial u\over\partial z}+{\partial p
\over\partial z}+{\partial S\over\partial z}+{\partial\over\partial z}
{{\bf b}^2\over 8\pi}=0 , &(2)\cr
&\biggl({\partial\over\partial t}+u{\partial\over\partial z}\biggr)\biggl(
{p\over\rho^\gamma}\biggr)+{\partial q\over\partial z}=0, &(3)\cr
&\rho{\partial {\bf v}\over \partial t}+\rho u{\partial {\bf v}\over\partial z}-
{B_0\over 4\pi}{\partial {\bf b}\over\partial z}=0 , 
 &(4)\cr
&{\partial {\bf b}\over\partial t}+u{\partial {\bf b}\over\partial z}
+{\bf b}{\partial u\over\partial z}
-B_0{\partial {\bf v}\over\partial z}=0 ,  &(5)\cr}
$$
where the velocity components are ${\bf v}={\bf V}_\bot , u=V_\|$ , the
magnetic field components are ${\bf b }={\bf B}_\bot , B_0=B_\| , \rho$ is
the mass density, $p$ is the pressure, $\gamma$ is the polytropic constant.
Also, $S$ and $q$ are the parallel momentum and heat fluxes respectively, i.e.
$S=-\rho\mu_\|(\partial u/\partial z)$ and 
$q=-n\chi_\|(\partial T/ \partial z)$, where $n=\rho/m_i$ and $T$ is the
temperature. There is no longitudinal perturbation of the magnetic field, as 
$ {\bf \nabla}\cdot{\bf B}=0$. Eqs.~(1)-(5) are the equations of continuity,
longitudinal momentum, energy conservation (with $q=0$ being the adiabatic
equation), transverse momentum and transverse flux, respectively.
Dissipation, as stated above, appears only via either momentum flux
($\mu\not=0, \chi=0$) or heat flux ($\mu=0, \chi\not=0$).
We expand the system of Eqs.~(1)-(5) in powers of $\epsilon=b/B_0 $. 
To  avoid secularities at third order, we assume multiple time-scale 
dependence, i.e.  $t_{(2n)}\sim t\epsilon^{2n}$ are the independent variables.

The zeroth-order solution of the system (1)-(5) is the equilibrium:
$\rho_0=const$ , 
 $V_0=0, {\bf B}_0=B_0\widehat{\bf e}_z$ , where $\widehat{\bf e}_z$ is the
unit vector in $z$ direction.
The first-order solution of Eqs.~(4) and (5) is the linearized Alfv\'en wave, i.e.
${\bf b}_1={\bf b}_1(z\pm v_A t_{(0)}, t_{(2)}), {\bf v}_1= (v_A/B_0){\bf b}_1 +
\widehat{\bf v}_1(t_{(2)})$ , where $v_A^2=B_0^2/4\pi\rho_0$ is the
Alfv\'en velocity.
Eqs.~(1)-(3) also have dissipative acoustic waves as a solution. As we 
consider
travelling hydromagnetic waves propagating to the right, we set 
$\rho_1=u_1=0$ and and choose a minus sign in the argument of ${\bf b}_1$.

At second order, we consider $\mu_\|\not=0$ and
$\chi_\|\not=0$ models, separately.

{\bf \item{a)} $\mu_\|\not=0$ model.}

Eqs.~(1)-(3) become, in second order:
$$
\eqalignno{
&
{\partial \rho_2\over \partial t_{(0)}}+\rho_0{\partial u_2 \over \partial z}=0,
& (6)\cr
&
\rho_0{\partial u_2 \over \partial t_{(0)}}+{\partial p_2 \over \partial z}
-\rho_0\mu_\|{\partial^2 u_2\over\partial z^2}
=-{\partial \over \partial z}{b_1^2 \over 8\pi} ,
& (7)\cr
&
{\partial \over \partial t_{(0)}}(p_2-c_s^2\rho_2)=0 ,
&(8)\cr}
$$  
where $c_s^2=\gamma p_0/\rho_0$ is the sound speed. From this system, 
excluding the free sound wave solution (as in first order), we have:
$$
\eqalignno{
&u_2={v_A \over \rho_0}\rho_2+\widehat{u}_2(t_{(2)}) , & (9)\cr
&p_2=v_A^2\rho_2+\mu_\|v_A{\partial\rho_2\over\partial z}-{b_1^2\over
8\pi} , & (10)\cr
&p_2=c_s^2\rho_2+\widehat{p}_2(t_{(2)}) . &(11)\cr}
$$  
Solving this system of equations, we, finally, arrive at the equation for the
second-order velocity perturbation:
$$
{u_2}_{(\mu)}={v_A^2 \over 2\mu_\|}e^{-z/L_\mu} \int^z e^{z'/L_\mu}
\left[{b_1^2(z')-\left<b_1^2(z')\right>\over B_0^2}\right] dz' ,
\eqno (12)
$$
where $\beta\equiv c_s^2/v_A^2$ and
$L_\mu=(\mu_\|/v_A)(1-\beta)^{-1}$. This is the
inhomogeneous solution of Eq.~(11), finite for all $-\infty<z<+\infty$.
The term $\left< b_1^2\right>$ 
appears in Eq.~(12) from density conservation: 
$d\left<\rho\right>/dt=0$. Note that it is the deviation from the mean
ponderomotive force which causes density bunching and thus wave
steepening. Thus $\left<\rho\right>=\rho_0$, so that 
$\left<u_2\right>=(v_A/\rho_0)\left<\rho_2\right>=0$.

{\bf \item {b)} $\chi_\|\not=0$ model.} 

Eqs.~(1)-(3) in this case are replaced by:
$$
\eqalignno{
&
{\partial \rho_2\over \partial t_{(0)}}+\rho_0{\partial u_2 \over \partial z}=0 ,
& (6)\cr
&
\rho_0{\partial u_2 \over \partial t_{(0)}}+{\partial p_2 \over \partial z}=
-{\partial \over \partial z}{b_1^2 \over 8\pi} ,
& (7')\cr
&
{\partial \over \partial t_{(0)}}(p_2-c_s^2\rho_2)=-{\partial q_2\over 
\partial z} ,
&(8')\cr}
$$  
The conductive heat flux can be represented as follows:
$$
q_2=-n_0 \chi_\|{\partial \over \partial z}T_2-n_1\chi_\|{\partial \over 
\partial z}T_1-n_2\chi_\|{\partial \over \partial z}T_0 .
\eqno (13)
$$
The second term here vanishes because $n_1=\rho_1/m_i=0$, as we set in 
the first-order equations. 
The third term vanishes also because we consider only a homogeneous
plasma temperature $T_0=const$. The second-order temperature pertubation
is, in turn, given by:
$$
T_2={p_2-T_0n_2 \over n_0}={p_2-v_t^2\rho_2 \over n_0} ,
\eqno (14)
$$
where $v_t^2=T_0/m_i$ is the thermal velocity of the particles. Now, from
Eqs.~(6),~(7'),~(8'), and using (13) and (14) we obtain:
 $$
\eqalignno{
&u_2={v_A \over \rho_0}\rho_2+\widehat{u}_2(t_{(2)}) , & (9)\cr
&p_2=v_A^2\rho_2-{b_1^2\over 8\pi}+\widehat{p}_2(t_{(2)}), & (10')\cr
&(p_2-c_s^2\rho_2)+{\chi_\| \over v_A}{\partial \over \partial z}
(p_2-{c_s^2 \over \gamma}\rho_2)=0 . &(11')\cr}
$$  
From these equations we obtain the second-order velocity perturbation for
the  $\chi_\|\not=0$ model as:
$$
\eqalignno{
{u_2}_{(\chi)}&={v_A \over 2(1-\beta/\gamma)}\biggl\{ {b_1^2-\left<b_1^2
\right> \over B_0^2} \cr
& +{v_A \over \chi_\|}\Bigl(1-{1-\beta \over 1-\beta/\gamma}
\Bigr) e^{-z/L_\chi}\int^ze^{z'/L_\chi} \left[{b_1^2(z')-\left<b_1^2(z')
\right> \over B_0^2}\right] dz'\biggr\} .
& (12')\cr}
$$
Here $L_\chi=(\chi_\|/v_A)(1-\beta/\gamma)/(1-\beta)$.
The velocities ${u_2}_{(\mu)}$ and ${u_2}_{(\chi)}$ are expressed in the frame 
co-moving with a wave (i.e. $z\to z-v_At_{(0)}$). 
The coefficient $\gamma=3$ is chosen to ensure energy
conservation~[17]. 

Eqs.~(4) and (5) are, at second order, equations for linear Alfv\'en waves.
One may easily
show that for all $t_{(2)}$ one may set ${\bf b}_2$ and ${\bf v}_2$ to zero [4].
At third order, upon substituting first- and second-order solutions into 
Eqs.~(4) 
and (5) one obtains the equation for $\partial {\bf b_1}/\partial t_{(2)}$:
$$
{\partial {\bf b}_1 \over \partial t_{(2)}}=-{1 \over 2}{\partial \over \partial z}
(u_2 {\bf b}_1) .
\eqno (15)
$$
For a dispersive term [6,24] one must invoke finite Larmor radius effects.
Since
Eq.~(15) describes the transverse components of ${\bf b}$, there is no 
influence of parallel dissipation
effects on the linear dispersive term. We thus add a general form of 
this term to obtain a DNLS-like magnetic perturbation envelope equation.  
Introducing $\tau=t_{(2)}$ and $\phi=({b_1}_x+i{b_1}_y)/B_0$  we have:
$$
{\partial \phi\over \partial\tau}+ {1 \over 2}{\partial \over \partial z}
(u_2 \phi)\mp i{v_A^2 \over 2\Omega_i}{\partial^2\phi \over\partial z^2}
=0 ,
\eqno (16)
$$
where $\Omega_i=eB_0/m_ic$ is the ion-cyclotron frequency, upper $(-)$
and lower $(+)$ signs on the third term refer to right and left elliptically
(circularly) polarized waves.
Eq.~(16) with (12) or (12') inserted for $u_1$ describes the coherent nonlinear 
dynamics of an Alfv\'en wave train subject to damping.
Damping results in the appearance of a new nonlocal 
integral contribution to the envelope equation. 
We define the integral dissipation operator as:
$$
{\cal J}_L[F](x)=e^{-x/L}\int^x e^{x'/L} F(x')dx' ,
\eqno (17)
$$
where $F$ is the arbitrary function the operator acts on and $L$ is a
characteristic length proportional to the dissipation coefficient
(i.e. $L\sim\mu_\|/[v_A(1-\beta)]$ or $L\sim \chi_\|/[v_A(1-\beta)]$ ).

From comparison of Eqs.~(12), (12') to the result of Ref.~[13] we conclude that
three-moment approximation ($\chi_\|\not=0$ model) of Landau damping
recovers the correct functional dependence $u_2\simeq c_1b_1^2+c_2{\cal L}
[b_1^2]$, where ${\cal L}$ is a nonlocal operator, and $c_1$ and $c_2$ are 
some
coefficients. However, the integral operator in this case (Eq.~(17)) is different 
from the
resonant particle operator of Ref.~[13] (see Eq.~(21) in the next section).
The two-moment approximation ($\mu_\|\not=0$ model) describes the
nonlinear physics incorrectly. Indeed, it lacks a free cubic nonlinear
contribution to the expression for $u_2$. This is in agreement with the
conclusions of Ref.~[17], which, stated simply, are the higher the moment 
approximation used, the better the description
of Landau damping in a fluid model which results. In particular, the linear 
response function of the three-moment fluid model
is much closer to the exact Maxwellian linear response function then that
of the two-moment fluid model. From now on we use the three-moment
fluid model ($\chi_\|\not = 0$).
\vskip 1.0cm
{\bf III. Coherent Nonlinear Alfv\'en Waves with Landau Damping }
\vskip 0.5cm
Using the $\chi\not=0$ model, 
kinetic effects are modeled by a longitudinal heat flux. In the linear closure
approximation, this flux is (in wave-number space [17]):
$$
q_{2_k}=-n_0\chi_1{\sqrt 2 v_t \over |k|}ikT_{2_k} .
\eqno (18)
$$
Here the temperature perturbation is given by Eq.~(14), and $\chi_1$ is a
dimentionless fit coefficient for the model. The choice $\chi_1=2/\sqrt \pi$
gives the best fit to linear Landau damping. By performing the inverse 
Fourier transform of
$q_{2_k}$, we obtain the real-space representation of $q_2$ as:
$$
\eqalignno{
q_2(z)&={1 \over \sqrt {2\pi}}\lim_{\delta\to 0}\int_{-\infty}^\infty
dk e^{-|k|\delta} e^{ikz}q_{2_k} \cr
&=-{n_0\chi_1\sqrt 2 v_t \over \pi}\int_0^\infty {T_2(z+z')-T_2(z-z') 
\over z'} ~dz'\cr
&=-{n_0\chi_1\sqrt 2 v_t \over \pi}\int_{-\infty}^\infty{{\cal P}\over z'-z}
T_2(z')dz' . &(19) \cr}
$$
Here ${\cal P}$ stands for the Cauchy principal value integral. We have also
added the factor of $\exp (-|k|\delta)$ to control the otherwise infinite 
integral. From comparison of this equation with Eq.~(13), we conclude that 
the effect of resonant particles in our theory is reduced to the 
following replacement:
$$
{\chi_\|\over v_A}{\partial \over\partial z}\to \widehat\chi_\|{\cal L} .
\eqno (20)
$$
We thus define here the resonant particle (nonlocal) integral operator:
$$
{\cal L}[F](x)={1\over\pi}\int_{-\infty}^\infty {{\cal P}\over x'-x}F(x')dx' 
\eqno (21)
$$
with the coefficient $\widehat \chi_\|=\chi_1\sqrt 2 v_t/v_A$ .

We now consider properties of the resonant particle operator in more 
detail. We need to know the inverse operator, ${\cal L}^{-1}$, so that 
${\cal L}^{-1}[{\cal L}]=1$. It is easy to find in $k$-space representation, i.e.
${\cal L}^{-1}{\cal L}={\cal L}_k^{-1}i{k/|k|}=
{\cal L}_k^{-1}i~ \hbox{sign}(k)=
1$, thus, ${\cal L}_k^{-1}=-i~ \hbox{sign}(k)$ and, returning to real space:
$$
{\cal L}^{-1}=-{\cal L} .
\eqno (22) 
$$
Formula (22) reflects the fact that Landau damping is time reversable, i.e. 
that a system returns to its initial state when time is reversed. There is no 
microscopic information loss in a Vlasov system, unlike in the case of 
collisional
damping. Writing an evolution equation in the form $(\partial/\partial t +
{\cal L})\psi=0$, we see that after time inversion ($t\to-t$),  
${\cal L}$ must satisfy Eq.~(22) to represent damping.
It can also be shown that the operator ${\cal L}$ commutes with any standard
differential or integral operator, i.e.
$$
{\eqalign{
&{\partial\over\partial z}({\cal L}[F(z)])={\cal L}\Bigl[{\partial\over\partial
 z}F(z)\Bigr]   , \cr
&\int ({\cal L}[F(z)])dz={\cal L}\Bigl[\int F(z)dz\Bigr]   , \cr
&{\cal L}[c]=0   , \cr
}}\eqno(23)
$$
where $c=const$.

Making a replacement, Eq.~(20), in the energy conservation equation~(11'), 
we have:
$$
\biggl( (1-\beta)+\Bigl(1-{\beta\over\gamma}\Bigr)\widehat\chi_\|
{\cal L}\biggr) \rho_2 = {\rho_0\over2}\Bigl(1+\widehat\chi_\|{\cal L}
\Bigr) {b_1^2-\left<b_1^2\right> \over B_0^2}  .
\eqno(24)
$$
This equation can be easily solved for $\rho_2$ using Eq.~(22).
The second-order velocity perturbation is then:
$$
u_2 = v_A \Bigl\{ M_1 \left|\Phi\right|^2 + M_2 {\cal L}
\left[ \left|\Phi\right|^2\right]\Bigr\} , 
\eqno(25)
$$
where $\left|\Phi\right|^2\equiv|\phi|^2-\left<|\phi|^2\right> 
=(b_1^2-\left<b_1^2\right>)/B_0^2$, and the factors $M_1$ and $M_2$ are:
$$
\eqalignno{
M^{(1)}_1&={1\over2}{(1-\beta)+\widehat\chi^2_\|(1-\beta/\gamma)\over
 (1-\beta)^2+\widehat\chi^2_\|(1-\beta/\gamma)^2 } , &(26a)\cr
M^{(1)}_2&=-{1\over2}\widehat\chi_\|\beta{\gamma-1\over\gamma}
{1\over (1-\beta)^2+\widehat\chi^2_\|(1-\beta/\gamma)^2 } ,&(26b) \cr}
$$
Here $\widehat\chi_\|=\chi_1 \sqrt{2\beta/\gamma},\  \gamma=3$
and $\chi_1=2/\sqrt\pi$ according to Ref.~[17]. The superscript (1) means
that the result refers to the one-fluid model.

Eq.~(16) together with Eq.~(25) recovers the modified derivative nonlinear
Schr\"odinger equation governing the envelope of nonlinear Alfv\'en
modes subject to Landau damping. Fig.~1 represents the dependence
of the factors  $M^{(1)}_1$ and $M^{(1)}_2$ 
on $1/\beta$ for comparison with the work by Spangler [13]. This case
corresponds to the $T_e/T_i=1$ case of Spangler. The coefficient $M_1$ of
Spangler is positive, always. The coefficient  $M_1^{(1)}$ in our
analysis changes sign at $\beta=1$, although where negative, its absolute
value is  small.
The disagreement is related to the lack of higher moments in the
closure scheme we use. The coefficients $M_2^{(1)}$ and $M_2$ of Spangler 
are negative and look alike. Analytical expressions for the coefficients in our
approach are much simpler than those obtained from kinetic calculations. 
The phenomenon which is described here by Eqs.~(25), (26) is typical
for other resonance phenomena. The sharp resonance without damping (here
the resonance between Alfv\'en and sound waves) is, of course, smoothed by
increasing damping. In other words, there is resonance broadening
caused by the interaction of  particles with a wave. In this respect, the
coupling to dissipation calculated here is the coherent analogue of nonlinear
ion Landau damping.

Until now we have considered only one-fluid magnetohydrodynamics. Based 
on this
model, we derived the Alfv\'en-sound wave resonance broadening in a
DNLS-like equation.
Astrohysical plasmas are essentially collisionless, i.e. the wave-length 
is much larger than mean free path, $\lambda\ll \ell_{{mfp}}$. Thus,  an
ion-acoustic mode replaces the sound wave in the collisionless regime.
Hence,  it is natural to adopt the perspective that 
the nonlinearity of finite-amplitude Alfv\'en waves arises from
an ion density perturbation, that is  from coupling of the ion-acoustic
and  Alfv\'en branches of the plasma oscillation. In other words, we must
consider electrostatic density perturbations instead of pure neutral density
perturbation as in a sound wave. The electric fields of the ion-acoustic wave
play a role in the dynamics of oblique nonlinear Alfv\'en waves, as will be
shown later.  To include the ion-acoustic
wave coupling effects instead of the gas-kinetic pressure perturbation, we
must use two-fluid dynamics for  both electron and ion fluids.
The longitudinal momentum equation, Eq.~(2), for ions in a collisionless 
plasma is:
$$
\rho{\partial u\over\partial t}+\rho u{\partial u\over\partial z}
+e{\partial \varphi\over\partial z}+{\partial\over\partial z}
{{\bf b}^2\over 8\pi}=0 ,
\eqno(27)
$$
where $\varphi$ is the electric potential. We neglect wave
damping by electrons, which is always much smaller than damping on ions,
i.e. ${\gamma_{{elect.}}/\gamma_{{ion}}}\simeq {m_e/m_i}$. 
Due to the quasineutrality of ion-acoustic waves, the first nonvanishing 
term of the expansion of the Boltzman distribution for electrons is:
$$
\widetilde n_i\simeq\widetilde n_e=n_0{e\varphi\over T_e} .
\eqno (28)
$$
Upon substituting this into Eq.~(27), and formally defining (for ions) $p^*_2=
n_2 T_{0e}=(T_e/T_i)p_2$, we can write Eq.~(7') as:
$$
\rho_0{\partial u_2\over\partial t_{(0)}}+{T_e\over T_i}{\partial
p_2\over\partial z}=-{\partial \over\partial z}{b_1^2\over 8\pi} .
\eqno (29)
$$
Multiplying Eq.~(8') by the factor $T_e/T_i$ and rewriting it with $p^*_2$
we have for this new pressure the same equation, with the 
$c^2_{s_i}\equiv(T_e/T_i)c^2_s$, - the speed of an ion-acoustic wave. 
We do the same with the closure equation (14).
Thus, all equations still apply for the new pressure $p^*_2$ when we 
substitute
$$
\beta \to {T_e\over T_i}\beta
\eqno (30)
$$
everywhere except in the $\widehat \chi_\|$ coefficient. There $\beta$ 
appears as a combination of the ion thermal velocity and the Alfv\'en 
velocity.

When $T_e/T_i\gg1$ an ion-acoustic wave experiences almost no damping
(neglecting the feeble damping by electrons). In the opposite case,
 $T_e/T_i\simeq1$, there is strong ion 
Landau damping. This is the case of the one-fluid calculation.
So, we must redefine the fit coefficient:
$$
\chi_1\to\chi_1 ~f\left({T_e\over T_i}\right)
\eqno(31)
$$
with some correction function $f(T_e/T_i)$ that reflects the behavior in two 
asymptotic regimes. This function must reflect the temperature dependence 
of the damping rate, and must equal unity at $T_e/T_i=1$. The damping 
rate of ion-acoustic waves in a Maxwellian plasma is well known [22]:
$$
{\gamma\over\omega}\simeq\biggl({\pi\over 8}\biggr)^{1/2}
\biggl({T_e\over T_i}\biggr)^{3/2}\exp\biggl\{-{T_e\over 2T_i}\biggr\} .
\eqno(32)
$$
Then, for  the correction function $f$ we obtain:
$$
f\Bigl({T_e\over T_i}\Bigr) = \biggl({T_e\over T_i}\biggr)^{3/2}
\exp\biggl\{-{T_e\over 2T_i}+{1\over 2}\biggr\} .
\eqno(33)
$$
Upon substitutions given by Eqs.~(30) and (31) in Eqs.~(26), we arrive at:
$$
\eqalignno{
M_1&={1\over2}{\left(1-{T_e\over T_i}\beta\right)
+\widehat\chi^2_\|\left(1-{T_e\over T_i}{\beta\over\gamma}\right)\over
 \left(1-{T_e\over T_i}\beta\right)^2+
\widehat\chi^2_\|\left(1-{T_e\over T_i}{\beta\over\gamma}\right)^2 } , 
&(34a)\cr
M_2&=-{1\over2}\widehat\chi_\|{T_e\over T_i}\beta
{\gamma-1\over\gamma}{1\over\left(1-{T_e\over T_i}\beta\right)^2+
\widehat\chi^2_\|\left(1-{T_e\over T_i}{\beta\over\gamma}\right)^2 } 
&(34b)\cr}
$$
with $\widehat\chi_\|=\chi_1 \sqrt{2\beta/\gamma}~ f(T_e/T_i)$.
Figs.~2a and 2b show the functions $M_1$ and $M_2$, respectively, plotted
 vs. $1/\beta$ for different electron-to-ion temperature ratios ($T_e/T_i = 
1, 3, 5, 10$). These figures closely resemble the graphs of Spangler [14] 
and Mj\o lhus and Wyller [6], however the expressions Eqs.~(34) are much 
simpler than
their counterparts in Refs.~[6,14], and easily amenable to detailed
calculation and evaluation,
etc.. The coefficient $M_2$ is always negative and has a  peak near 
$\beta_0=T_i/T_e$. There is  slight difference when $T_e\simeq T_i$. 
The coefficient $M_1$ should be positive in this case, for almost all values 
of $\beta$. This difference can result from insufficient accuracy of the
three-moment approximation ($\chi_\|\not=0$) when Landau damping is
very strong. To get a better fit one should take higher-moment 
approximations. Of course, for large amplitude Alfv\'en waves the 
assumption of a Maxwellian plasma fails, as well.
\vskip 1.0cm
{\bf IV. Slightly Oblique Nonlinear Alfv\'en Waves with Landau Damping 
and Finite Larmor Radius Effects}
\vskip 0.5cm
In the previous sections we considered Alfv\'en waves propagating strictly
parallel to the magnetic field lines. Here we try to generalize our approach 
to waves propagating at a small angle to the magnetic field. It has been shown
[12, 23-25]
that dissipationless oblique nonlinear Alfv\'en waves are governed by the
DNLS equation, as well. We investigate here the influence of kinetic effects on
wave dynamics. To include kinetic effects in the fluid model of the DNLS 
we use a two-fluid gyrofluid model~[18,19]. Here we construct
the simplest model which includes kinetic effects such as finite
gyro-frequency,
Landau damping, etc. We omit gyro-viscous corrections to the Reynolds 
stress. These corrections are small for the slightly oblique propagation case
in straight field geometry, and can be incorporated (approximately) into 
the viscous model considered in Section II.

Typically, the gyrofluid equations are written in components projected onto 
the
magnetic field direction, $z'$, and perpendicular to it. It is more convenient to
us to work in  the frame of a wave propagating in the $z$ direction, which 
makes some angle $\Theta\ll 1$ relative to the $z'$ direction. 
We choose the $y$ direction
coincident with $y'$. We leave variables `unprimed' when measured in the 
frame of a wave, and use primes for  variables measured in the frame of
the ambient magnetic field. We consider contributions of kinetic effects 
separately for every equation of the system~(1)-(5).
The equation for the transverse flux is obtained from the Maxwell equations
$\nabla\times{\bf E}=(-1/c)\partial{\bf B}/\partial t$, $\nabla\times
{\bf B}=(4\pi/c){\bf J}$ and the Ohm's law. We write the generalized Ohm's
law as follows:
$$
{\bf E}=-{1 \over c}[{\bf v}\times{\bf B}]+{m_i \over ec\rho}[{\bf J}\times
{\bf B}]-{m_i\over 2e\rho}\nabla p+\nabla\varphi_{el} .
\eqno (35)
$$
There are no contributions from the last two terms. There is no contribution
from the term $\nabla\varphi\times\nabla\rho$ because $\varphi$ is the
first-order perturbation, and we have set $\rho_0=const, \rho_1=0$. The
first term has already been accounted for. The second term leads to an
additional term on the right-hand-side of Eq.~(5). We write it in components
as follows:
$$
\eqalignno{
&{\partial b_x\over\partial t}+{\partial\over\partial z}(ub_x)-{B_0}_z
{\partial v_x\over\partial z}+{B_0}_x{\partial u\over\partial z}=-{v_A^2
\over\Omega_i}\left({B_0}_z{\partial^2\over\partial z^2}b_x\right)
  ,&(36a)\cr
&{\partial b_y\over\partial t}+{\partial\over\partial z}(ub_y)-{B_0}_z
{\partial v_y\over\partial z}={v_A^2\over\Omega_i}\left({B_0}_z
{\partial^2\over\partial z^2}b_y\right)
  .&(36b)\cr}
$$
The terms on the right-hand-side give rise to a linear dispersion term in the
DNLS. This term describes dispersion due to finite ion Larmor radius.
In gyrofluid models there is no correction to the transverse momentum
equation, Eq.~(4). We write it with the substitution of $B_0$ by ${B_0}_z$,
as is obvious for oblique propagation:
$$
\rho{\partial {\bf v}\over\partial t}+\rho u{\partial{\bf v}\over\partial z}=
{{B_0}_z\over 4\pi}{\partial{\bf b}\over\partial z} .
\eqno (37)
$$
In the gyrofluid equations [18,19] there are
corrections from both ${\bf E}\times{\bf B}$ velocity and an effect of
gyro-averaging of fields over the Larmor orbit. The ${\bf E}\times{\bf B}$ 
drift velocity ${\bf v}_E=c[{\bf E}\times{\bf B}]/B^2$ enters the 
gyrofluid analogs of Eqs.~(1)-(3) in the combination ${\nabla\cdot(u'
\widehat {\bf e}_{z'}+{\bf v}'_E)}$, where $\widehat {\bf e}_{z'}$ - is the unit
sector in $z'$ direction. Upon substituting the electric field from Eq.~(35)
we obtain the following. The first term contributes to the velocity $u'$, giving
rise to $\nabla\cdot(u'\widehat {\bf e}_{z'}+{\bf v}'_\bot)=\nabla\cdot
{\bf v}=\partial u/\partial z$, i.e. recovering the usual velocity divergence
term. The second term does not make a contribution because $(4\pi/c){\bf J}=
\nabla\times{\bf B}=\nabla\times\widetilde {\bf b}_\bot$ and, thus, 
${\bf J}_\bot\equiv 0$. The last two terms are both gradient terms. For a 
plane wave the 
gradient is reduced to $\nabla=\widehat {\bf e}_z(\partial/\partial z)$. 
So, both these terms contribute to 
${\bf v}_E$ like $[\widehat {\bf e}_z
\times\widehat {\bf e}_{z'}]\sim\widehat {\bf e}_y$. Upon substituting into 
the divergence term of Eqs.~(1)-(3), it vanishes. Thus, we conclude that for
a plane single coherent wave, the ${\bf E}\times{\bf B}$ drift does not make 
a significant contribution to the nonlinear Alfv\'en wave dynamics.

Another effect we should consider is the gyro-averaging of electric and
magnetic fields over the Larmor orbits of particles. This effect enters
the equations via a differential operator which approaches
unity when the ion Larmor radius
vanishes. Since it acts on fields only and because we construct the model
which must reduce to MHD in the case of zero ion Larmor radius, finite
gyro-radius corrections appear only in Eq.~(2), leaving Eqs.~(1) and (3)
unchanged. In the case of oblique propagation, Eqs.~(1)-(3) are:
$$
\eqalignno{
&{\partial\rho\over\partial t}+{\partial\over\partial z}(\rho u) =0 , &(38)\cr
&\rho{\partial u\over\partial t}+\rho u{\partial u\over\partial z}=
-\left<J_0\right>\biggl\{{\partial p^* \over\partial z}
+{\partial\over\partial z} {b_x^2+b_y^2\over 8\pi}+{{B_0}_x\over 4\pi}
{\partial b_x\over\partial z}\biggr\}=0 , &(39)\cr
&\biggl({\partial\over\partial t}+u{\partial\over\partial z}\biggr)\biggl(
{p^*\over\rho^\gamma}\biggr)+{T_e\over T_i}{\partial q\over\partial z}=0
 . &(40)\cr}
$$
Here we have already used the quasineutrality condition together with
Eq.~(28) for the electric potential perturbation. This appears on the
right-hand-side of Eq.~(39) as a gradient of $p^*$. This accounts for the 
charge separation in an ion-acoustic wave, and the 
subsequent gyro-averaging of the electric field associated with the electron
responce. Eq.~(40) is a trivial rewriting of Eq.~(3) for the
quantity $p^*$.

The operator $J_0$, which carries out the gyro-averaging operation, is a
linear operator. It is simply a Bessel function represented in Fourier space:
$$
\eqalignno{
J_0\left({k'_\bot v'_\bot\over\Omega_i}\right)
&=\int_0^{2\pi}d\vartheta\exp\left\{i{k'_\bot v'_\bot\over\Omega_i}
\cos{\vartheta}\right\}
=\sum^\infty_{n=0}{1\over(n!)^2}\left({{v'}_\bot^2 \over2\Omega_i^2}
\right)^n{\nabla'}_\bot^{2n} , &(41)\cr}
$$
where $\nabla'_\bot$ is the gradient in the plane perpendicular to $z'$. 
Going to the frame of the wave, it
contains a longitudinal projection equal to $\sin{\Theta}(\partial/\partial z)$.
All other components vanish for a plane wave in a homogeneous plasma. The
operator $\left<J_0\right>$ is the operator $J_0$ averaged over the
(Maxwellian) particle distribution function. There are different ways [19] to
approximate $\left<J_0\right>$. We choose one of the more simple
approximations:
$$
\left<J_0\right>\cong\Gamma_0^{1/2}\simeq
1-{\rho_i^2\over2}{\nabla'}_\bot^2 + ...\simeq 1-{\rho_i^2\over2}
\sin^2{\Theta} {\partial^2\over\partial z^2} ,
\eqno(42)
$$
where $\rho_i=v_t/\Omega_i$ is the ion Larmor radius. We neglect, as usual,
terms of higher order in $\rho_i^2$.

Equations (36)-(40) replace Eqs.~(1)-(5) for the case of an obliquely
propagating wave subject to Landau damping and including finite Larmor
radius effects. The procedure for derivation of the DNLS-like
equation describing nonlinear, obliquely propagating Alfv\'en waves is similar
to that explained in Section II for parallel-propagating waves, but requires a
bit more algebra. From Eqs.~(36) and (37) we finally arrive at the
DNLS-like equation, Eq.~(16), with a new field:
$$
\psi={{b_1}_x+i {b_1}_y+{B_0}_x\over {B_0}_z}
\eqno(43)
$$
instead of $\phi=({b_1}_x+i {b_1}_y)/B_0$. One can mention that in terms of
the field $\phi$ the DNLS for oblique waves is more complicated and contains
Korteweg-de Vries nonlinearities. These terms are proportional to ${B_0}_x$
and, thus, disappear in the case of strictly parallel propagation.

The velocity perturbation $u_2$ is found from the second-order expansion of
Eqs.~(38)-(40). After one integration these equations can be written in the 
frame moving with the wave as:
$$
\eqalignno{
&v_A\rho_0u_2=\Gamma_0^{1/2}p^*_2
+\Gamma_0^{1/2}{{b_1}_x^2+{b_1}_y^2\over 8\pi}
+\Gamma_0^{1/2}{{B_0}_x {b_1}_x^2\over4\pi} , &(44)\cr
&\Bigl(p_2^*-{c_{s_i}^2\rho_0\over v_A}u_2\Bigr)
+\widehat\chi_\|{\cal L}\Bigl[p_2^*
-{c_{s_i}^2\rho_0\over\gamma v_A}u_2\Bigr]=0 . &(45)\cr}
$$
Upon acting on the second equation with the operator $\Gamma_0^{1/2}$ and
substituting $\Gamma_0^{1/2}p^*_2$ from the first, we obtain the
equation for $u_2$ as follows:
$$
\left\{\left(1-\beta^*\Gamma_0^{1/2}\right)+\widehat\chi_\|{\cal L}
\left(1-{\beta^*\over\gamma}\Gamma_0^{1/2}\right)\right\}[u_2]=
{v_A\over2}\left\{1+\widehat\chi_\|{\cal L}\right\}\Gamma_0^{1/2}
\left[\left|\Psi\right|^2\right] , 
\eqno(46)
$$
where $\beta^*\equiv(T_e/T_i)\beta$ and $\left|\Psi\right|^2=|\psi|^2-
\left<|\psi|^2\right>$.
We solve this equation using the inverse operator (22) and
the commutation relations (23). Using approximation for the operator 
$\Gamma_0^{1/2}$ given by Eq.~(42), we finally obtain the following 
equation for $u_2$:
$$
\left(1-\Lambda^2{\partial^2\over\partial z^2}\right) u_2=v_A
\left\{\left(M_1-N_1{\partial^2\over\partial z^2}\right)
+{\cal L}\left(M_2-N_2{\partial^2\over\partial z^2}\right)\right\}
\left[\left|\Psi\right|^2\right] , 
\eqno(47)
$$
where
$$
\eqalignno{
&\Lambda^2=-2\eta^2_\|\beta^*{(1-\beta^*)+
{\widehat\chi^2_\|\over\gamma}(1-\beta^*/\gamma)\over
(1-\beta^*)^2+\widehat\chi^2_\|(1-\beta^*/\gamma)^2 } ,&(48a)\cr
&N_1={1\over2}\eta^2_\|{(1-2\beta^*)+\widehat\chi^2_\|(1-2\beta^*/
\gamma)\over(1-\beta^*)^2+\widehat\chi^2_\|(1-\beta^*/\gamma)^2 } ,
&(48b)\cr
&N_2=-\eta^2_\|\widehat\chi_\|\beta^*{\gamma-1\over\gamma}{1\over
(1-\beta^*)^2+\widehat\chi^2_\|(1-\beta^*/\gamma)^2 } 
.&(48c)\cr}
$$
Here we used the notation $\eta^2_\|\equiv(\rho_i^2/2)\sin^2{\Theta}$, and
$M_1$ and $M_2$ are given by Eqs.~(34). The solution of this equation
can be easily obtained in terms of the integral dissipation operator 
${\cal J}$ given by Eq.~(17). We thus have:
$$
u_2=v_A\left\{\left(M_1+M_2{\cal L}\right){1\over 2\Lambda}\left(
{\cal J}_\Lambda-{\cal J}_{-\Lambda}\right)+\left(N_1+N_2{\cal L}\right)
{1\over\Lambda^2}\right\} \left[\left|\Psi\right|^2\right] .
\eqno(49)
$$
This is the inhomogeneos solution valid for all values of $\Lambda^2$. 
There is the finite homogenious solution when $\Lambda^2$ is negative, i.e. 
$\beta^*<1$:
$$
{u_2}_{hom}={u_2}_{in}~\sin(z/|\Lambda|) ,
\eqno(49')
$$
 where $ {u_2}_{in}$ is the constant that can be obtained from initial
conditions. This solution represents,
in a sense, a free modulation wave, i.e. a travelling modulation of the envelope
of the nonlinear Alfv\'en wave train with amplitude set by initial 
conditions.
When $\Lambda^2$ becomes positive, this solution of the 
homogenious equation diverges at infinity like ~$\sinh(z/\Lambda)$.

The dissipation operator combination can be written as:
$$
\eqalignno{
{1\over 2\Lambda}\left({\cal J}_\Lambda-{\cal J}_{-\Lambda}\right)[F]
&={1\over 2\Lambda}
\left[\int^ze^{(z'-z)/\Lambda}F(z')dz'-\int^ze^{-(z'-z)/\Lambda}
F(z')dz'\right] \cr
&~~~~~\cr
&= {\cases{
\displaystyle{
{1\over\Lambda}\int^z\sinh\bigl[(z'-z)/\Lambda\bigr]F(z')dz'}
 , & if $\Lambda^2 > 0$ ; \cr
~~~~& ~~~~\cr
\displaystyle{
{-1\over |\Lambda|}\int^z\sin\bigl[(z'-z)/|\Lambda|\bigr]F(z')dz'}
 , & if $\Lambda^2 < 0$ . \cr} }
&(50)\cr}
$$
It is interesting to note that the kernel of this integral is an antisymmetric 
function
of $z'-z$, as in the resonant particle integral, but is not singular at $z'=z$.
Thus, the main contribution to this integral is not from the instantaneous 
position $z$, unlike the resonant particle operator. 
It is also interesting that this operator does not correspond to 
pure dissipation, because the kernel of the operator ${\cal J}_L$
itself consists of both symmetric and antisymmetric parts.

When $\Lambda^2\simeq0$ we can write the approximate solution of
Eq.~(47). Using an iterative method, we obtain:
$$
u_2=v_A\left\{\left(M_1+M_2{\cal L}\right)+\left[\left(\Lambda^2M_1-N_1
\right)+\left(\Lambda^2M_2-N_2\right){\cal L}\right]
{\partial^2\over\partial z^2}\right\}\left[\left|\Psi\right|^2\right] .
\eqno(51)
$$
Fig.~3 represents the dependence of $\Lambda^2/\eta_\|^2$ vs. 
$1/\beta$ for the same $(T_e/T_i)$ values as in Figs.~2a and 2b. This quantity 
changes sign at $\beta\simeq\beta_0$.  It defines the particular structure of 
the finite Larmor radius integral operator given by Eq.~(50).
The analogous plots of $N_1/\eta_\|^2$ and
$N_2/\eta_\|^2$ are shown in Figs.~4a and 4b respectively. 
These coefficients are negative (there is a region
where $N_1$ is positive, but very small), peaked near $\beta_0$ and look
similar to graphs of the coefficient $M_2$.
\vskip 1.0cm
{\bf V. Modulation Instability }
\vskip 0.5cm
It has been demonstrated~[9,26,27] that modulation instability leads to the
destruction of wave trains and the production of solitons, transfer of energy 
from large-scale
wave-modes to (damped) small-scale modes, the appearance of shock waves 
and shocklets in the vicinity of bow-shocks of planets, etc.. As was already
mentioned, waves described by the DNLS can be modulationly unstable
depending on the sense of their polarization. Left-hand polarized waves are
unstable when the cubic nonlinearity coefficient is positive, and right-hand
polarized waves are unstable when it is negative.
In this Section we investigate the modulation instability of dissipative wave
packets described by the modified DNLS, Eqs.~(16) and (12'), and compare 
the result with that of the DNLS with the resonant particle term. We write 
the general equation as follows:
$$
{\partial\phi\over\partial\tau}+{\partial\over\partial z}\Bigl[\phi\bigl(Q_1
\left(|\phi|^2-\left<|\phi|^2\right>\right)+Q_2{\cal J}_L\left[|\phi|^2-
\left<|\phi|^2\right>\right]\bigr)\Bigr]-i\mu{\partial^2\over\partial 
z^2}\phi=0 ,
\eqno (52)
$$
where
$$
\eqalignno{
&Q_1={v_A\over 4}{1\over 1-\beta/\gamma} ,\qquad 
Q_2=-\beta{v_A^2\over 4\chi_\|}{\gamma-1\over\gamma}
{1\over (1-\beta/\gamma)^2}
&(53)\cr}
$$
and $\mu=v_A^2/2\Omega_i$. We consider only left-hand polarized waves
(i.e. with a minus sign on the third term) which are modulationally stable 
when  $Q_1>0$ in the dissipationless limit, $Q_2=0$.

Our approach follows that of Mio {\sl et.al.}~[28] and Spangler~[14].
Expressing the solution of Eq.~(52) in the form:
$$
\phi=ue^{i\vartheta} ,
\eqno (54)
$$
where $u(\tau,z)$ and $\vartheta(\tau,z)$ are real functions and
substituting this into Eq.~(52), we obtain two equations for the real and 
imaginary parts, respectively:
$$
\eqalignno{
&\dot u+Q_13u^2u_z-Q_1\left<u^2\right>u_z+Q_2u_z{\cal J}_L\left[u^2-
\left<u^2\right>\right] \cr
&\qquad+Q_2u\left({\cal J}_L\left[u^2-\left<u^2\right>\right]
\right)_z+2\mu\vartheta_zu_z+\mu\vartheta_{zz}u=0 ,
&(55a)\cr
&\dot\vartheta+Q_1\vartheta_zu^3-Q_1\vartheta_z\left<u^2\right>u
+Q_2\vartheta_zu{\cal J}_L\left[u^2-\left<u^2\right>\right]-\mu u_{zz}+
\mu\vartheta_z^2u=0
 ,&(55b)\cr}
$$
where  $\dot u\equiv\partial u/\partial t$, $u_z\equiv\partial u/\partial z$,
 and similarly for $\vartheta$ and ${\cal J}_L$. Using the solutions $U_0$
and $\Theta_0$, we superimprose small-amplitude and phase modulation
such as:
$$
\eqalignno{
&u=U_0+\epsilon\widetilde u ,
\qquad \vartheta=\Theta_0+\epsilon\widetilde\vartheta ,
&(56)\cr}
$$
where $\epsilon\ll 1, U_0=const, \dot\Theta_0=-\omega_0, (\Theta_0)_z=
k_0$, where, in turn, $\omega_0$ and $k_0$ are the frequency and
wavenumber of high-amplitude Alfv\'en wave. Eqs.~(55) with $u$ and
$\vartheta$ of the form~(56) yields (in zeroth order of $\epsilon$) the
dispersion relation:
$$
\omega_0=\mu k_0^2 .
\eqno (57)
$$
Note that no nonlinearity enters the dispersion relation when obtained in
the frame moving with a wave.

In first order the linearized Eqs.~(55) become:
$$ 
\eqalignno{
&\dot{\widetilde u}+\widetilde u_z(2Q_1U_0^2+2\mu k_0)+2Q_2U_0^2(
{\cal J}_L[\widetilde u])_z+\widetilde\vartheta_{zz}\mu U_0=0 
 ,&(58a)\cr
&\dot{\widetilde\vartheta}+2\widetilde\vartheta_z\mu k_0+2\widetilde u
Q_1k_0U_0-\widetilde u_{zz}{\mu\over U_0}+2Q_2k_0U_0{\cal J}_L
[\widetilde u]=0
 ,&(58b)\cr}
$$
To go further, we need to know the Fourier representation of the operator 
${\cal J}_L$. We write:
$$
\eqalignno{
{\cal J}_L[\widetilde u]&=\int_{-\infty}^0 e^{\zeta/L}\widetilde u(z+\zeta)d
\zeta=\int_{-\infty}^0 e^{\zeta/L}\left(\int_{-\infty}^{\infty}dk~ u_k e^{ik
(z+\zeta)}\right)d\zeta \cr
&=\int_{-\infty}^{\infty}dk~ e^{ikz}{u_k\over ik+1/L} .
&(59)\cr}
$$
Taking the Fourier transformation of Eqs.~(58), with $\widetilde u,
\widetilde\vartheta\sim\exp(ikz-i\omega\tau)$, we obtain:
$$
\eqalignno{
&(-i\omega+A_r+iA_i)u_{k,\omega}-C_r\vartheta_{k,\omega}=0
 ,&(60a)\cr
&(B_r-iB_i)u_{k,\omega}+(-i\omega+iC_i)\vartheta_{k,\omega}=0
 .&(60b)\cr}
$$
The coefficients are:
$$
\eqalign{
&A_r=2Q_2U_0^2L^2k^2 ,\quad A_i=k\left(2Q_1U^2+2\mu k_0+{2Q_2U_0
^2L\over 1+k^2L^2}\right) ,\cr
&B_r=2Q_1U_0k_0+{2Q_2U_0Lk_0\over 1+k^2L^2}+k^2{\mu\over U_0}
 ,\quad B_i={2Q_2U_0L^2k_0k\over 1+k^2L^2} ,\cr
&C_r=k^2\mu U_0 ,\quad C_i=2\mu kk_0 .\cr}
\eqno (61)
$$
Instability appears when the 
imaginary part of the compex frequency $\omega$
is positive. When $Q_2$ vanishes there is no instability, so typically $\gamma
\ll\omega_r$. Substituting the frequency $\omega=\omega_r+i\gamma$,
we may write an expression for the growth rate $\gamma$ as:
$$
\gamma\simeq\mu k_0{|Q_2|\over |Q_1|}{k^2L^2\over 1+k^2L^2} .
\eqno (62)
$$
It is noticable that the growth rate in this case does not depend on the
amplitude of the nonlinear Alfv\'en wave, unlike the resonant particle case.
In astrophysical plasma $\beta\simeq 1$, so that $k^2L^2\gg1$, i.e. the
modulation length is much smaller the characteristic length of dissipation
$L$. In this case:
$$
\gamma\sim {v_A^3k_0\beta\over \Omega_i\chi_\|} .
\eqno (62')
$$
In the opposite case $k^2L^2\ll1$:
$$
\gamma\sim {v_A\chi_\|k_0\over l^2_m\Omega_i}
\beta{(1-\beta/\gamma)^2\over (1-\beta)^2} ,
\eqno (62'')
$$
where $l_m=1/k$ is the modulation length. 
One can see that those  waves which are modulationally {\sl stable} in the 
{\sl dissipationless}
limit become {\sl unstable} when {\sl collisional dissipation increases}. 
The growth rate
of modulation instability is proportional to the ratio of the coefficients of the 
nonlocal and the cubic nonlinearity terms. 
When the nonlocal term coefficient is zero, we
obtain from the system~(60) the known criterion of modulation
stability which
depends on the sense of polarization of the wave~[10,14].

Modulation stability analysis of the modified DNLS with the resonant
particle integral operator has been implemented in Refs.~[14,29]. The 
growth rate for this case is:
$$
\gamma\sim{\mu^2 k_0^2|k|\over B_0^2U_0^2M_1}
\left({|M_2|\over M_1}\right) .
\eqno (63)
$$
Using Eqs.~(34) for the case of low beta, $\beta\ll1$, we obtain the scaling:
$$
\gamma\sim\widehat\chi_\|\beta{T_e\over T_i}\left({v_A^4k_0^2\over
\Omega_i^2|\widetilde{\bf b}|^2 l_m}\right) .
\eqno(63')
$$
Modulation instability now appears  for waves which are stable in the absence
of resonant particle effects. The growth rate again is
proportional to the nonlocal-to-nonlinear term coefficient ratio. Roughly,
this growth rate is $M_2/M_1$ times smaller than the modulation instability
growth rate predicted by the simple DNLS.
\vskip 1.0cm
{\bf VI. Conclusions }
\vskip 0.5cm
In this paper we have considered the influence of kinetic effects on nonlinear
Alfv\'en wave dynamics. The effects under consideration are i) the dissipative
(collisional) longitudinal viscous and thermal fluxes,
ii) resonant particle effects, i.e. Landau damping, iii) the
${\bf E}\times {\bf B}$ drift and finite Larmor radius effects such as those
associated with the 
gyro-averaging of fields over the Larmor orbits of particles. All calculations
have been based on fluid moment and gyrofluid models with Landau 
damping modelled by additional dissipation-like terms. The results obtained 
are given below.
\item{a.}
Dissipative (collisional) longitudinal viscosity, $\mu_\|$, and
thermoconductivity, $\chi_\|$, give rise to an additional, new integral
nonlinear term in the evolution equation for  the nonlinear Alfv\'en waves, 
and also resolve the $(1-\beta)^{-1}$ singularity of 
the derivative nonlinear term in the MHD model.
The dissipative integral operator is different from the integral operator
representing resonant particle effects, and is given by Eq.~(17). 
\item{b.}
Modulationally stable
dissipationless Alfv\'en waves become unstable when dissipation is included.
The growth rate is proportional to the ratio of the coefficients of the integral 
and nonlinear terms. This is similar to the case of the modified DNLS 
with Landau damping, studied in Ref.~[14].
\item{c.} 
Using the three-moment fluid equations [17], we derived the modified DNLS 
including Landau damping effects. Since in this model dissipation
is no longer an algebraic constant, but an integral operator given by
Eq.~(21), we obtained exactly the same functional form of the equation as that 
obtained from lengthy, full kinetic calculations [6,14]. However, the 
coefficient of both the nonlinear and resonant particle operator (nonlocal) 
terms obtained here are much simpler,
thus facilitating clear physical interpretation and further analysis. 
Particularly, the phenomenon under study is just 
the broadening of the resonance between acoustic and Alfv\'enic 
branches due to the resonant particle interaction with a sound wave. 
\item{d.}
In the collisionless regime, there is coupling of Alfv\'en and
ion-acoustic waves, instead of coupling of Alfv\'en and sound waves.
The resonant particle effect is then strong Landau damping 
of the ion-acoustic wave when $T_e/T_i\simeq1$. We obtained the 
temperature ratio dependent coefficients $M_1$ and $M_2$ of the nonlinear 
and integral terms, respecitvely. These coefficients are much simpler than
those obtained full Vlasov model  [14]. However, they display
similar qualitative dependence on the  $\beta$ and $T_e/T_i$ parameters, 
Figs.~2a, 2b.
\item{e.}
In the case of obliquely propagating waves, other kinetic effects may
be relevant. It was shown that even though the ${\bf E}\times {\bf B}$ drift
velocity is not zero (electric field is generated from charge separation in an
ion-acoustic wave), it does not contribute to the dynamic wave
equation. Another effect which does contribute is a gyro-averaging of electric 
and magnetic fields acting on
a particle over a Larmor orbit of this particle. This effect depends on the
angle between the wave propagation direction and the ambient magnetic
field direction as ~$\sin^2{\Theta}$, and vanishes for a strictly 
parallel-propagating wave. We studied this effect using a gyrofluid model
constructed from gyrofluid equations [18,19]. This effect results in additional
terms in the amplitude evolution equation. These terms are expressed in 
terms of the dissipation integral operator, given in Eq.~(17). Hence, this 
effect  can be viewed as some specific nonlinear dissipation 
(similar to collisional dissipation) of an obliquely propagating
wave due to finite Larmor radius. 

We have shown in this paper that MHD models with linear kinetic 
corrections  provide correct quantitative description of effects
such as the dynamics of high-amplitude Alfv\'en waves. It is significant
that the expressions obtained are, nevertheless, much simpler 
than those obtained from kinetic (Vlasov) calculations,
thus facilitating further analysis and numerical calculations, and allowing 
very clear physical interpretation. There are some
points unresolved in this work. First, when the amplitude of a wave is
high, there are particles trapped in the wave. 
These particles traverse the regions of lower 
field and bounce between the regions of higher field. This
trapped motion in a beat wave results in amplitude oscillation of this wave
at the particle bounce frequency, similar to the nonlinear Landau damping 
process.
Obviously, such phenomena cannot be represented by a theory based on an
assumption that the distribution function deviates slightly from a local
Maxwellian. Obviously, the physical process of trapping will lead to plateau
formation and other modification of Maxwellian structure. Indeed, waves
with $b_\bot/B_0\sim1$ will have large trapping width and undoubtedly
lead to significant distortion of the distribution function. It is interesting to
note that such distortions could potentially {\sl mitigate} the effects of Landau
damping, via local flattening of $\left< f\right>$ at the resonant velocity 
(for $\beta\sim1$).
Second, the gyrofluid model of this paper is the simplest possible. 
It is important to consider full set of nonlinear gyrofluid equations to 
include gyro-viscosity, finite Larmor radius corrections to the Reynolds 
stress, etc., which are relevant for the case of oblique and near-perpendicular 
propagation. 

As noted above, the kinetic effects which govern the dynamics of a
collisionless plasma may significantly alter the pictire of nonlinear Alfv\'en
wave dynamics built upon the MHD plasma model. Inclusion of
wave-particle resonance effects can significantly alter predictions for
modulational instability (i.e. left vs. right circular polarization dependence of
growth rate upon parameters such as $T_e/T_i, \ \beta$, etc.). Also, the
nonlocal structure of the envelope equation which arises from the effects of
parallel streaming, will likely result in departure from the traditional
paradigm of  collisionless shocks as solitons formed by the
competition between nonlinear steepening and dispersion. In particular, a
new time scale, namely the ion transit time through the envelope modulation,
enters along with the steepening and dispersion rates. Strong ion heating will
occur, as well. Thus, collisionless shock structure may be smoothed or exhibit
secondary temporal oscillations. These speculations may be easily addressed
by studies of numerical solution of the (tractable) envelope equations derived
in this paper. The result of these numerical studies will be published in
Part~II of this series.

Finally, it should be mentioned that an improved understanding of nonlinear
Alfv\'en dynamics in a compressible plasma may have application in contexts
other than collisionless shocks in the solar wind plasma. First, a significant
fraction of the interstellar medium [30] is hot, collisionless compressible
plasma. Thus, the problems of the galactic dynamo and interstellar
turbulence should be approached in the context of a collisionless,
compressible plasma model. In particular, collisionless dissipation via ion
heating is a natural mechanism for controlling the growth of small scale
magnetic energy which has been shown to inhibit the mean field dynamo in
purely incompressible MHD theories [31,32]. Similarly, the processes of wave 
steepening and shock or soliton formation can strongly affect the parallel
dynamics of Alfv\'enic turbulence in the interstellar medium. 
Such turbulence is thought to be related to interstellar scintillations. 
These issues will be addressed in future publications.

\vskip 1.0cm
{\bf Acknowledgements }
\vskip 0.5cm
We are grateful to V.I.~Shevchenko, V.D.~Shapiro and S.K. Ride for 
discussions and to E.~Mj\o lhus for his comments.
One of us (M.M.) is also grateful to S.R. Spangler for his interest
in this work.

This research was supported by U.S. Department of Energy, Grant No.
DE-FG03-88ER53275 and National Aeronautics and Space 
Administration Grant No. UT-A:NAGW-2418.
\vfill
\eject
{\bf References }
\vskip 0.5cm
\item{1.} B.T.~Tsurutani, E.J.~Smith, {\sl Geophys. Res. Lett.}, {\bf 13}, 
263 (1986).
\item{2.} B.T.~Tsurutani, P.~Rodriguez, {\sl J. Geophys. Res.}, {\bf 86}, 
4319 (1981).
\item{3.} R.Z.~Sagdeev, C.F.~Kennel, {\sl Scientific American}, No.~4, 106
(1991).
\item{4.} R.H.~Cohen, R.M.~Kulsrud, {\sl Phys. Fluids}, {\bf 17}, 2215
(1974).
\item{5.} A.~Rogister, {\sl Phys. Fluids}, {\bf 14}, 2733 (1971).
\item{6.} E.~Mj\o lhus, J.~Wyller, {\sl J. Plasma Phys.}, {\bf 40}, 299 (1988).
\item{7.} E.~Mj\o lhus, {\sl J. Plasma Phys.}, {\bf 19}, 437 (1978).
\item{8.} S.R.~Spangler, J.P.~Sheerin, G.L.~Payne,  {\sl Phys. Fluids}, 
{\bf 28}, 104 (1985).
\item{9.} S.P.~Dawson, C.F.~Fontan, {\sl Phys. Fluids}, {\bf 31}, 83 (1988).
\item{10.} M.~Longtin, B.U.\"O.~Sonnerup, {\sl J. Geophys. Res.}, {\bf 91},
6816 (1986).
\item{11.} H.K.~Wong, M.L.~Goldstein, {\sl J. Geophys. Res.}, {\bf 91},
5671 (1986).
\item{12.} E.~Mj\o lhus, J.~Wyller, {\sl Phys. Scr.}, {\bf 33}, 442 (1986).
\item{13.} S.R.~Spangler, {\sl Phys. Fluids B}, {\bf 1}, 1738 (1989).
\item{14.} S.R.~Spangler, {\sl Phys. Fluids B}, {\bf 2}, 407 (1989).
\item{15.} G.S.~Lee, P.H.~Diamond, {\sl Phys. Fluids}, {\bf 29}, 3291
(1986).
\item{16.} R.E.~Waltz,  {\sl Phys. Fluids}, {\bf 31}, 1963 (1988).
\item{17.} G.W.~Hammett, F.W.~Perkins, {\sl Phys. Rev. Lett.}, {\bf 64},
3019 (1990).
\item{18.} C.L.~Hedrick, J.-N.~Leboeuf, {\sl Phys. Fluids B}, {\bf 4}, 3915
(1992).
\item{19.} W.~Dorland, G.W.~Hammett, {\sl Phys. Fluids B}, {\bf 5}, 812
(1993).
\item{20.} S.E.~Parker, W.~Dorland, R.A.~Santoro, M.A.~Beer, Q.P.~Liu,
W.W.~Lee, G.W.~Hammet, {\sl Phys. Plasmas}, {\bf 1}, 1461 (1994).
\item{21.} A.~Brizard, {\sl Phys. Fluids B}, {\bf 4}, 1213 (1992).
\item{22.} Rus.: L.A.~Artsimovich, R.Z.~Sagdeev, {\sl Fizika plazmy dlya
fizikov (Plasma physics for physicists)}, p.~75, Moskva, Atomizdat 1979. 
\item{23.} C.F.~Kennel, B.~Buti, T.~Hada, R.~Pellat, {\sl Phys. Fluids}, 
{\bf 30}, 1949 (1988).
\item{24.} S.R.~Spangler, B.B.~Plapp, {\sl Phys. Fluids B}, {\bf 4}, 3356
(1992).
\item{25.} T.~Hada, C.F.~Kennel, B.~Buti, {\sl J. Geophys. Res.}, {\bf 94}, 65
(1989).
\item{26.} S.~Ghosh, K.~Papadopoulos, {\sl Phys. Fluids}, {\bf 30}, 1371
(1987).
\item{27.} S.R.~Spangler, {\sl Phys. Fluids}, {\bf 29}, 2535 (1986).
\item{28.} K.~Mio, T.~Ogino, K.~Minami, S.~Takeda, {\sl J. Phys. Soc. Jpn.},
{\bf 41}, 667 (1976).
\item{29.} T.~Fl\aa, 
E.~Mj\o lhus, J.~Wyller, {\sl Phys. Scr.}, {\bf 40}, 219 (1989).
\item{30.} L.~Spitzer, {\sl Diffusive Matter in Space}, New York, Interscience
Publishers 1968.
\item{31.} A.V.~Gruzinov, P.H.~Diamond, {\sl Phys. Rev. Lett.}, {\bf 72},
1651 (1994).
\item{32.} A.V.~Gruzinov, P.H.~Diamond, {\sl Phys. Plasmas}, {\bf 2}, 1941
(1995).
\vfill
\eject
~\vskip1.5in\hskip-1in
   \includegraphics{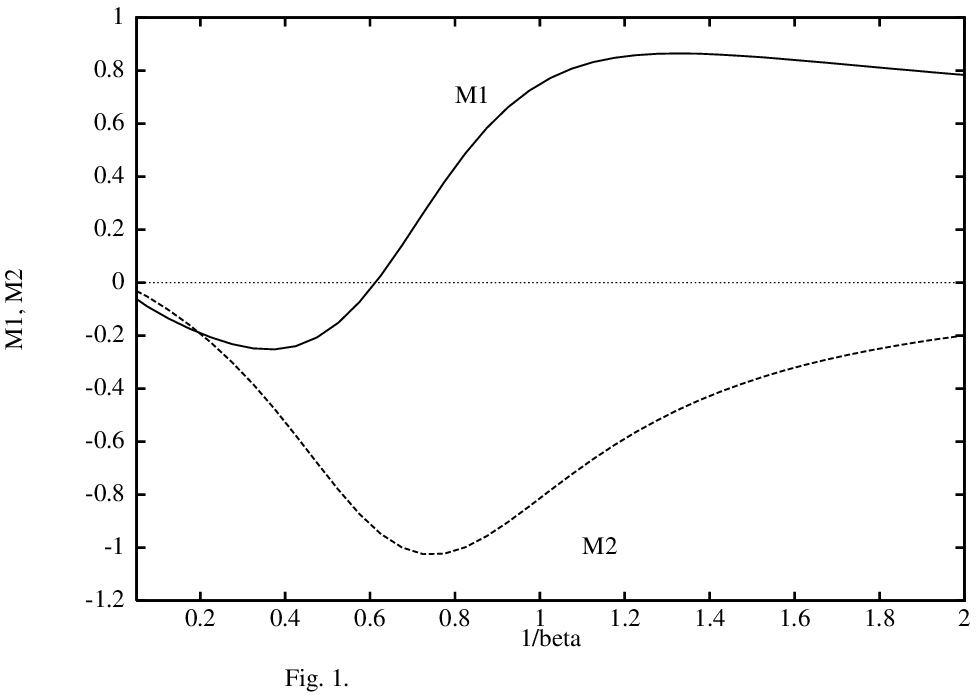}
\vskip2.5in
{\bf Fig.~1.} The coefficients of the nonlinear and integral terms, 
$M^{(1)}_1$ and $M^{(1)}_2$,  in the one-fluid model vs. $1/\beta$.
\vfill
\eject
~\vskip0.5in\hskip-1in
   \includegraphics{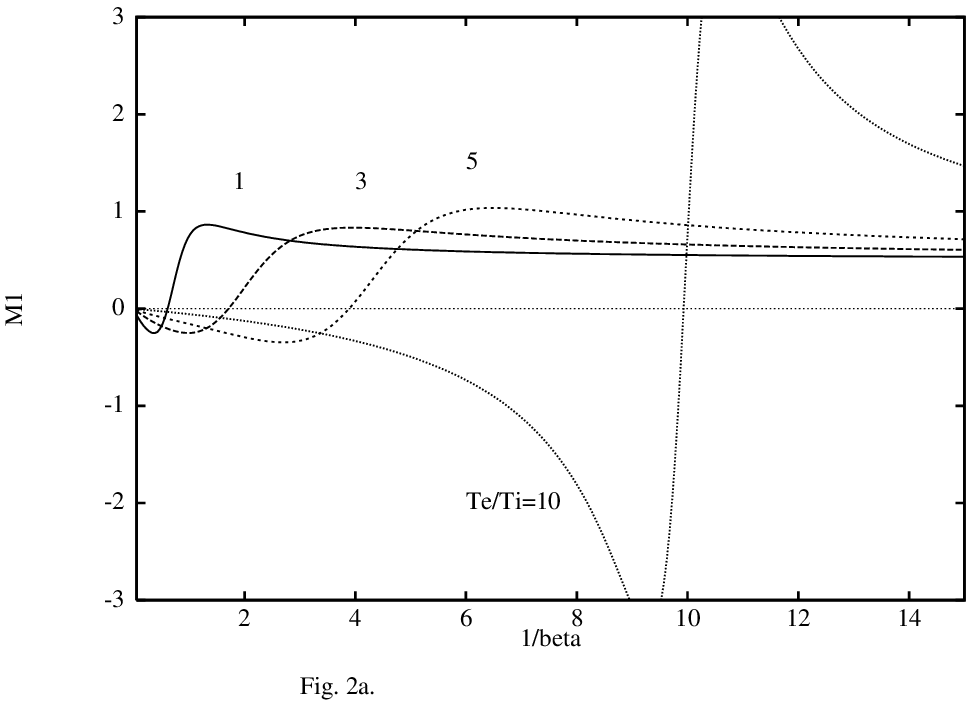}
\vskip2.8in\hskip-1in
   \includegraphics{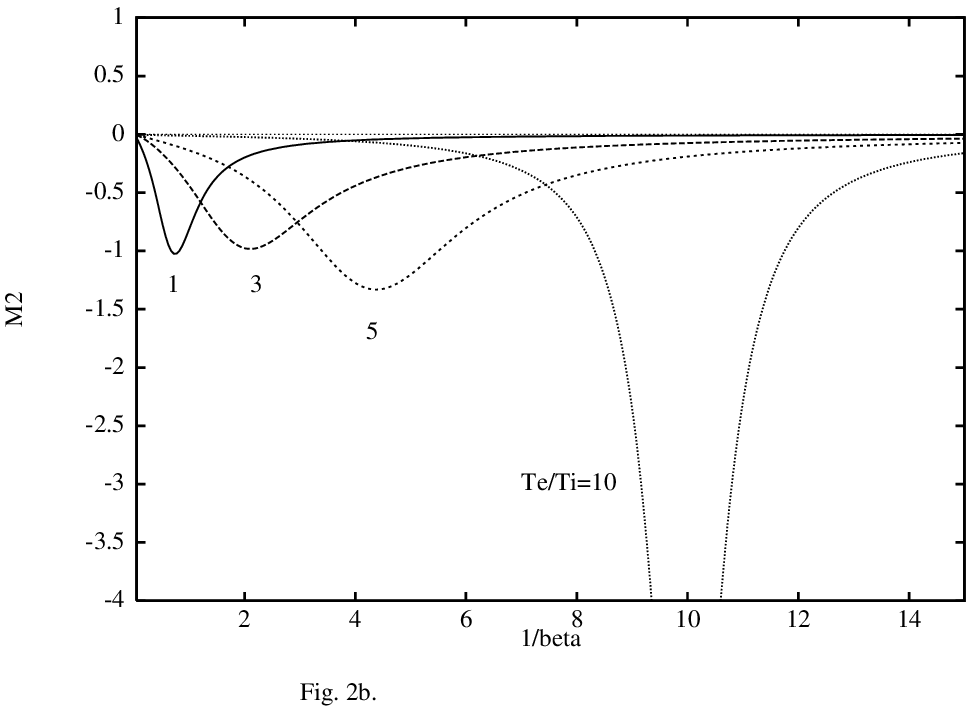}
\vskip2.5in
{\bf Figs.~2a, 2b.} The same coefficients as on Fig.~1 in the two-fluid model
vs. $1/\beta$ for four $T_e/T_i$ temperature ratios ($T_e/T_i =1, 3, 5, 10$).
\vfill
\eject
~\vskip1.5in\hskip-1in
   \includegraphics{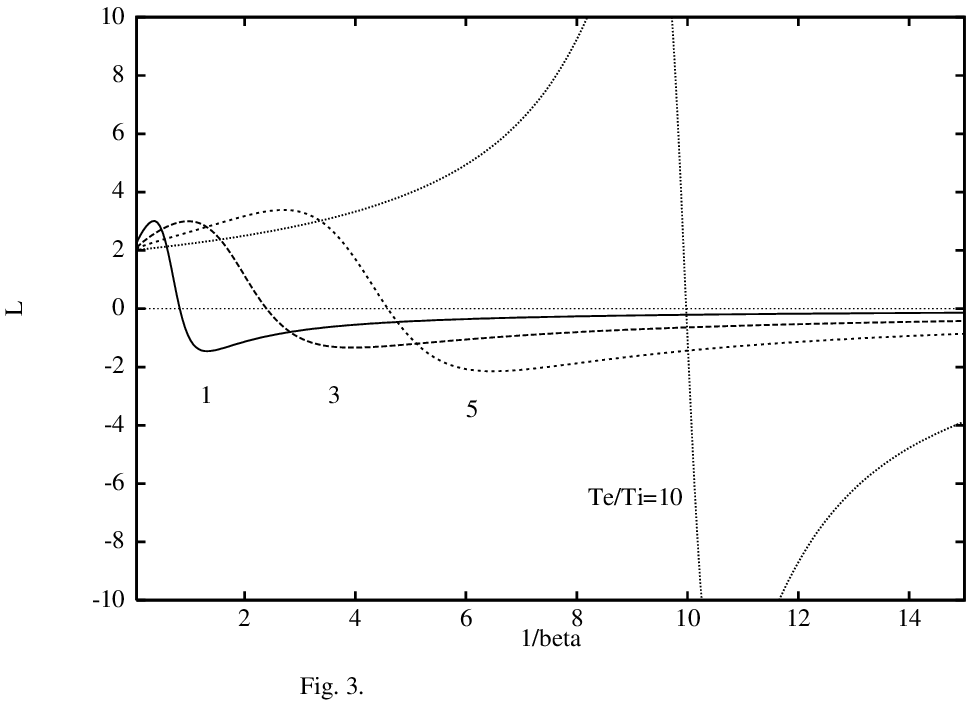}
\vskip2.5in
{\bf Fig.~3.} Same dependence as on Figs.~2a, 2b for the normalized quantity
$\Lambda^2/\eta_\|^2$.
\vfill
\eject
~\vskip0.5in\hskip-1in
   \includegraphics{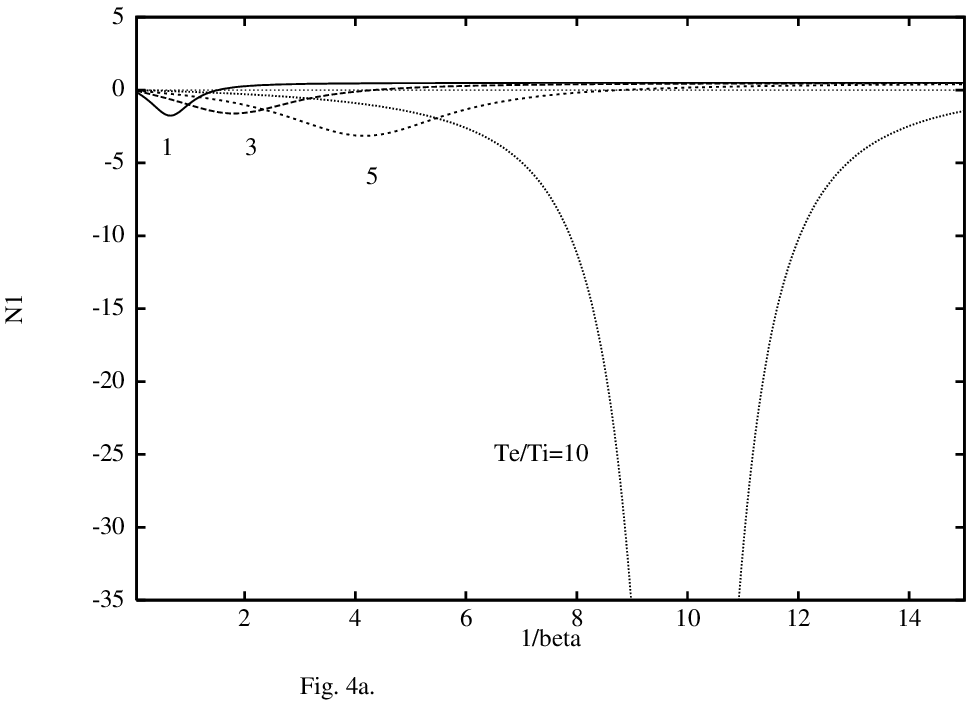}
\vskip2.8in\hskip-1in
   \includegraphics{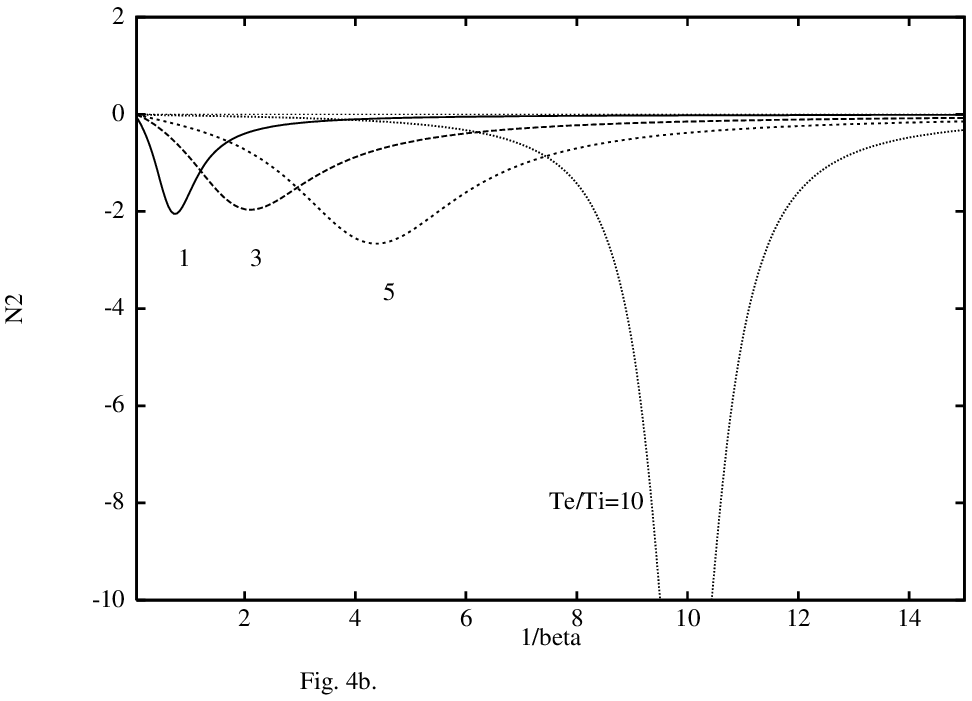}
\vskip2.5in
{\bf Figs.~4a, 4b.} Same as on Figs.~2a, 2b for the normalized coefficients 
$N_1/\eta_\|^2$ and $N_2/\eta_\|^2$.
\vfill
\eject
\end